\documentclass[11pt]{article}
\usepackage[sc]{mathpazo} 
\usepackage{fullpage}
\usepackage[authoryear,sectionbib,sort]{natbib}
\usepackage[utf8]{inputenc}
\usepackage{lineno}

\usepackage[usenames, dvipsnames]{color}
\usepackage{xcolor,color}
\usepackage{graphicx}
\graphicspath{{figures/}} 
\usepackage{amsmath,amssymb,amsthm,amsopn,mathrsfs}
\usepackage{caption}
\usepackage{subcaption}
\usepackage[toc,page]{appendix}
\usepackage{csvsimple}
\usepackage{setspace}
\definecolor{navy}{rgb}{0,0,0.502}
\newcommand*\patchAmsMathEnvironmentForLineno[1]{%
  \expandafter\let\csname old#1\expandafter\endcsname\csname #1\endcsname
  \expandafter\let\csname oldend#1\expandafter\endcsname\csname end#1\endcsname
  \renewenvironment{#1}%
     {\linenomath\csname old#1\endcsname}%
     {\csname oldend#1\endcsname\endlinenomath}}%
\newcommand*\patchBothAmsMathEnvironmentsForLineno[1]{%
  \patchAmsMathEnvironmentForLineno{#1}%
  \patchAmsMathEnvironmentForLineno{#1*}}%
\AtBeginDocument{%
\patchBothAmsMathEnvironmentsForLineno{equation}%
\patchBothAmsMathEnvironmentsForLineno{align}%
\patchBothAmsMathEnvironmentsForLineno{flalign}%
\patchBothAmsMathEnvironmentsForLineno{alignat}%
\patchBothAmsMathEnvironmentsForLineno{gather}%
\patchBothAmsMathEnvironmentsForLineno{multline}%
}

\title{Estimating global species richness using symbolic data meta-analysis}

\author{H. Lin$^{\ast,\dag}$,
	M. J. Caley$^{\ddag}$ and
	S. A. Sisson$^{\ast}$}

\date{}

\begin{document}
	
	\maketitle
	
	\noindent{}$\ast$ --  School of Mathematics and Statistics, University of New South Wales, Sydney, 2052, Australia;
	
	\noindent{}$\ddag$ --  School of Mathematical Sciences, Queensland University of Technology, Brisbane, Qld, 4001, Australia; 
	

	\noindent{}$\dag$ -- Corresponding author; E-mail: huan.lin@unsw.edu.au.
	
	
	

	 
	
	
	
	
	
	\linenumbers{}
	\modulolinenumbers[3]
	
	\newpage{}
	
	\section*{Abstract}
	
	Global species richness is a key biodiversity metric. Concerns continue to grow over its decline due to overexploitation and habitat destruction by humans. Despite recent efforts to estimate global species richness, the resulting estimates have been highly uncertain and often logically inconsistent. Estimates lower down either the taxonomic or geographic hierarchies are often larger than those above. Further, these estimates have been typically represented in a wide variety of forms, including intervals $(a,b)$, point estimates with no uncertainty,  and point estimates with either symmetrical or asymmetrical bounds, making it difficult to combine information across different estimates. Here, we develop a Bayesian hierarchical approach to estimate the global species richness that combines 45 estimates from published studies. The data mix of intervals and point estimates are reconciled using techniques from symbolic data analysis.  This approach allows us to recover interval estimates at each species level, even when data are partially or wholly unobserved, while respecting logical constraints, and to determine the effects of estimation on the whole hierarchy of obtaining future estimates for particular taxa at various levels in the hierarchy. Our estimate of global species richness from this meta analysis of 22.3 (11.2, 36.3) million eukaryotic species is substantially greater than some recent estimates. While our estimates combine information from many different studies using a variety of estimation methods, it is not our intention that these estimates be treated as consensus values, but rather a joint estimate based on prior information currently available that will and should be updated as new information becomes available. A measure of success of this method is that this estimate will change as new information becomes available and eventually stabilise. The uncertainties reported here should guide where additional research resources are directed to improve these estimates. It should no longer be sufficient to simply erect a new method of estimation and claim its primacy without careful justification and progressive validation into the future.
	
	\newpage{}
	
	\section*{Introduction}
	\label{introduction}
	
	
	Knowing the total number of species on Earth - or at least having a good approximation – has been a key but elusive goal for ecologists for many decades (\citealt{caley2014global}). Without an adequate baseline against which future changes in biodiversity can be compared, it will not be possible to know with any certainty what and how much biodiversity has been lost anytime in the future, the success or failure of conservation and recovery efforts, or indeed, even whether all species can be named before they become extinct (\citealt{costello2013can}).  It is also well accepted that global biodiversity is threatened by myriad agents (\citealt{hunter2007human}, \citealt{pimm2014biodiversity}), and that its accelerating loss (\citealt{ceballos2015accelerated}) will be associated with degraded ecosystem services (\citealt{benayas2009enhancement}, \citealt{mooney2009biodiversity}). Despite this urgency to understand the world’s biological resources and what could be lost, little progress has been made in achieving logically consistent estimates across taxa and ecological realms and estimates that converge through time both in terms of the most likely estimates and increasing confidence around these estimates (\citealt{caley2014global}). Recently, however, some evidence has begun to emerge that at least for terrestrial arthropods and some subtaxa thereof, global species richness estimates have begun to narrow (\citealt{stork2015new}). Nonetheless, while this narrowing of the ranges of these estimates is encouraging, substantial further narrowing is still needed because confidence in these species richness estimates still varies by more than 1 million to many millions of species depending on the group being considered (\citealt{stork2015new}).
	Moreover, this quest for better estimates of species richness has been dominated by the sequential development of estimation methods each considered in isolation. By considering these estimates in isolation from each other, information is lost that, if combined, might be able to be harnessed to achieve better estimates. For example, where a species richness estimate for a realm (e.g. all marine species: \citealt{may1992many}) is lower than an estimate for a marine habitat (e.g. coral reefs: \citealt{fisher2015species})  a source of uncertainty has been clearly identified and this information may be able to be leveraged in aid of a better estimate if this sort of information from separate studies can be combined in meaningful ways. Here we attempt to incorporate information across studies that estimate global species richness or components thereof,
	adopting a Bayesian hierarchical model to estimate species richness into logically consistent estimates across all species. Our approach not only enforces logical consistency of estimates across the hierarchy, but also improves estimation accuracy by sharing of 
	information across taxa.\\\\
	For our analysis, we used previously published estimates of the total number of species worldwide, as well as estimates of the number of species from various subcategories, including coral reefs, the marine environment, beetles, insects and terrestrial arthropods (Table \ref{Table:dataset}). Although there is a wealth of published species richness estimates at regional and subregional scales, we restrict our analyses here to global estimates because without robust estimates of the spatial turnover of species (beta-diversity) between locations that are not yet widely available, there is the danger of counting the same species multiple times, and thereby, inflating global estimates \cite{may1992bottoms}. The global estimates  data we used are
	 recorded in three forms. Some (17 estimates) are point-estimates, $x$,  representing experts' most informed best estimate of the number of species based on a detailed analysis or study. Some (19 estimates) are recorded as intervals $(a,b)$ representing a plausible range of species counts.  The remainder (9 estimates) provide both a point estimate and an interval, which may represent symmetric or asymmetric uncertainty. 
	One approach to combining these data into a joint analysis is to convert the interval-valued estimates to a single value, e.g. $x=(a+b)/2$, which would then permit a simple analysis on the means. However, taking the midpoint assumes that the uncertainty around this midpoint is symmetrical, and analysis of these single points ignores the often considerable uncertainty contained in the full interval.
	\cite{caley2014global} followed this approach, and also fitted independent models to the upper and lower interval endpoints to determine whether the species richness estimates converged over time.
However, assuming independence between these three models could produce self-inconsistent results whereby the lower interval boundary could exceed the upper boundary, and as before, information may be lost.\\
	As an alternative, we model these data, both single-values and intervals, using techniques from symbolic data analysis (see e.g. \citealt{billarddiday2007} for a comprehensive review).
	In this setting, the univariate interval $(a,b)\subseteq\mathbb{R}$ is mapped to the bivariate random vector $(m, \log r)^\top\in\mathbb{R}^2$, where $m=(a+b)/2$ represents the interval mid-point, and $r=(b-a)$ is the interval range (\citealt{brito2012modelling}). 
	Performing a statistical analysis on this bivariate vector is equivalent to analysing the univariate random interval $(a,b)$ (\citealt{zhang+s16}).
	Translating the intervals to midpoint and log range makes no assumption that the distribution within each interval is symmetric -- rather it is just a convenient reparameterisation.  Single valued estimates $x$ can be also directly expressed in this bivariate vector form, where the midpoint is given by the expert's best guess $m=x$, and the range $r$ (which describes the uncertainty on the estimate) is simply unobserved, so that $(m,\log r)^\top = (x, \mathrm{NA})^\top$.  Given this reparameterisation into a common data format, these bivariate random vectors can then be modelled via a Bayesian hierarchical model (e.g. \citealt{gelmanbook}, chapter 5).\\
	This model enforces logically consistent hierarchical relationships among different species categories, that for example, the number of insects plus the number of other arthropods must sum to the total number of arthropods. It can also produce estimates of species categories that have not yet been observed or recorded, and can provide estimates of the missing ranges $r$ for the point-estimate only data, thereby allowing the unobserved full interval to be predicted from the posterior.
	It additionally permits determination of the effects on model species richness estimates when including a new measurement in a particular category, thereby allowing an assessment of where best to focus future species richness estimation efforts, and provides a way to update global estimates as new estimates within the hierarchy become available.
%


\section*{Methods}
\label{methods}
The data comprise 45 previous estimates of species richness obtained from the literature (Table \ref{Table:dataset}), of which 42 were previously analysed by \cite{caley2014global}. Each paper provides a species estimate in one or more of 7 categories: coral reefs (4 estimates), marine species (8), terrestrial species (1), arthropods (10), insects (12), beetles (1) and global species estimates (9). These estimates come in either interval form $(a,b)$ (19 cases), point estimate form $x$ (17 cases) and both interval and point estimate form $(a,x,b)$ (9 cases). In the latter case, the best guess point estimate $x$ exactly coincides with the midpoint of the interval in  4 cases (so that $x=(a+b)/2$), but is different in the remaining 5 cases.
As described in the Introduction, we re-parametrise each species richness estimate into the bivariate vector $(m,\log r)^\top$ describing interval midpoint and log range. That is, where interval estimates $(a,b)$ or $(a,x,b)$ are available, these are expressed as $m=(a+b)/2$ and $r=(b-a)$, and where only a point estimate $x$ is available, then $m=x$, and $\log r$=NA is a missing value.

Note that in the 5 cases where a point estimate and interval are both available, but where the point estimate is not the midpoint of the interval, we ignore the point estimate, and still express the $m$ as the midpoint of $(a,b)$. While this potentially loses information about the possible asymmetric nature of the interval estimates in these 5 cases, we do this for two reasons. Firstly, modelling the midpoints as $m=(a+b)/2$ makes no assumptions of the distribution of the expert's estimate within the interval. This could be symmetric or asymmetric. Rather, it simply states that the midpoint of the interval itself is $m=(a+b)/2$. So treating the estimates $(a,x,b)$ in this way means that the midpoint $m$ has the same interpretation as the interval-only estimates $(a,b)$. By implication, this means that we are stating that for  point estimates $m=x$ only, that the point estimate is the centre of the unobserved interval, which then becomes a model assumption. Secondly, there are very few (5) abundance estimates with a point estimate that is different from the interval mid-point. While it is possible to construct an asymmetric model for a 3-dimensional reparameterisation of the trivariate vector $(a,x,b)^\top$ (e.g. \citealt{le2011likelihood} propose a way of modelling asymmetric interval-valued data assuming a triangular distribution), this would result in a large number of missing values for the rest of the dataset, far more than could be handled with confidence. It is therefore more realistic for the present analysis to restrict attention to bivariate modelling. Trivariate modeling will be more informative in the future as more $(a, x, b)^\top$ format data become available. \\
\begin{figure}[tb]
	\centering
	\includegraphics[width=13cm]{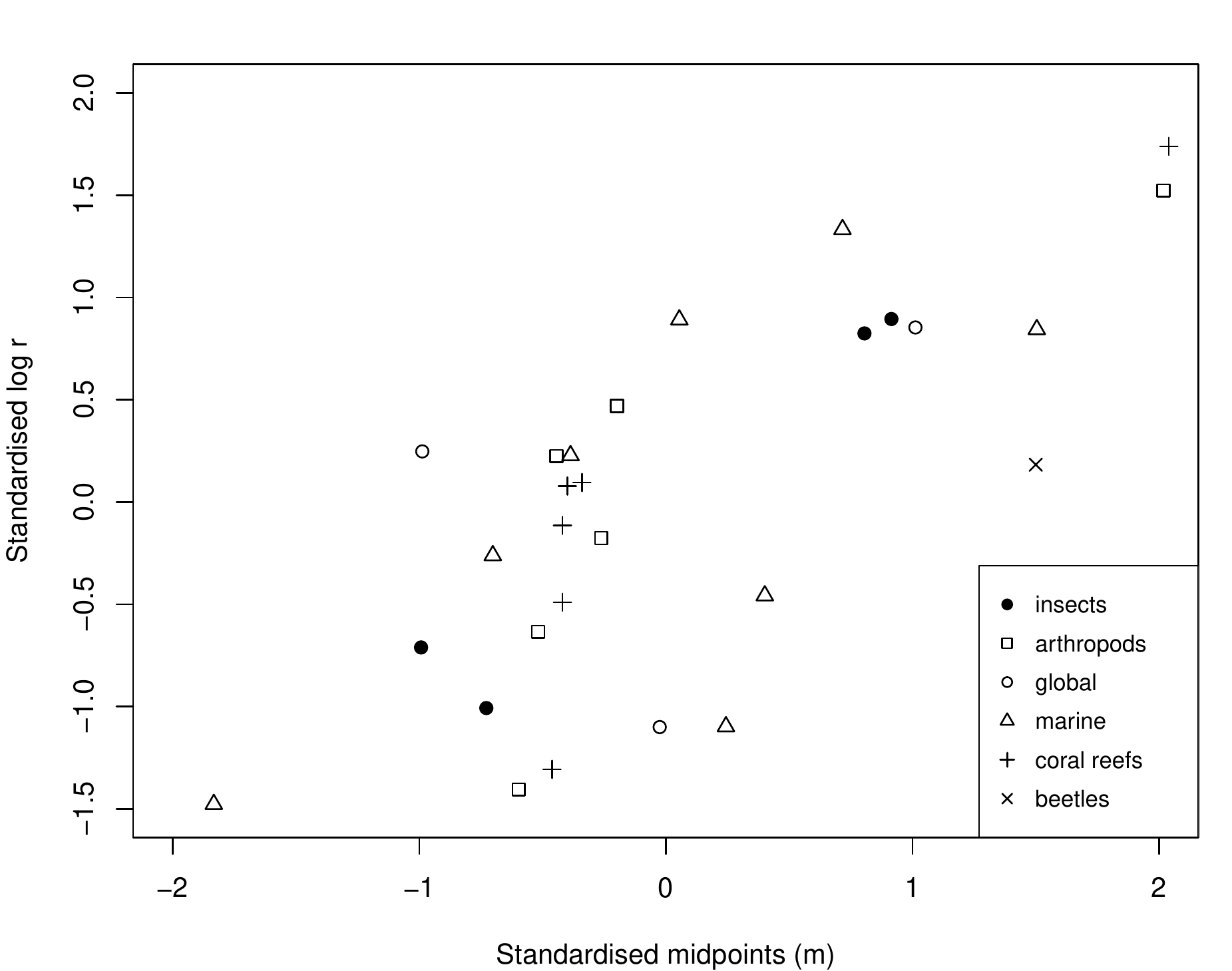}
	\caption{\label{fig:m-versus-logr}\small Scatterplot of standardised midpoints ($m$) versus standardised log range ($\log r$) for the 28 observed intervals. Point types indicate species category. 
	}
\end{figure}
For observed data within a given category $j\in\mathcal{C}$ with $\mathcal{C}=\{$global, other-global, marine, other-marine, arthropods, other-arthropods, coral-reefs, insects, other-insects, beetles$\}$ (Fig. \ref{fig:m-versus-logr}),  we can then model the derived $(m,\log r)^\top$ via an appropriate statistical model. A positive association between $m$ and $\log r$ is expected given that these data are species counts and the variability of count data increases with the number being counted. Visually, a bivariate Gaussian distribution could credibly represent these data well, albeit with different location and scale parameters for each category. That is, for each species category $j$, with observed data $(m_{ij},\log r_{ij})^\top$ for $i=1,\ldots,n_j$, we suppose that 

\begin{equation}
\label{eqn:model}
(m_{ij},\log r_{ij})^\top\sim N_2(\mu_j,\Sigma_j)
\end{equation}
where
\[
\mu_j=(\mu_{mj},\mu_{rj})^\top
\qquad\mbox{and}\qquad
\Sigma_j=\begin{bmatrix}
\sigma^2_{mj} & \rho\sigma_{mj}\sigma_{rj}  \\
\rho\sigma_{mj}\sigma_{rj} & \sigma^2_{rj} \\
\end{bmatrix}.
\]
Note that we specify the correlation $\rho$ between $m_{ij}$ and $\log r_{ij}$ to be the same across all species categories $j$. This decision is based on visual inspection of Figure \ref{fig:m-versus-logr} (with allowance for small sample sizes), in which the linear dependence appears similar across all categories, and the not unreasonable assumption that any species counting process is similar for all categories. The resulting advantage is that the correlation $\rho$ can be estimated using the observed data from all categories, and that this provides a way for the model to share information between categories. This can be particularly useful when estimating missing $\log r$ values for single valued data $(m,\mathrm{NA})^\top$.\\
\begin{figure}[tb]
	\centering
	\includegraphics[width=13cm]{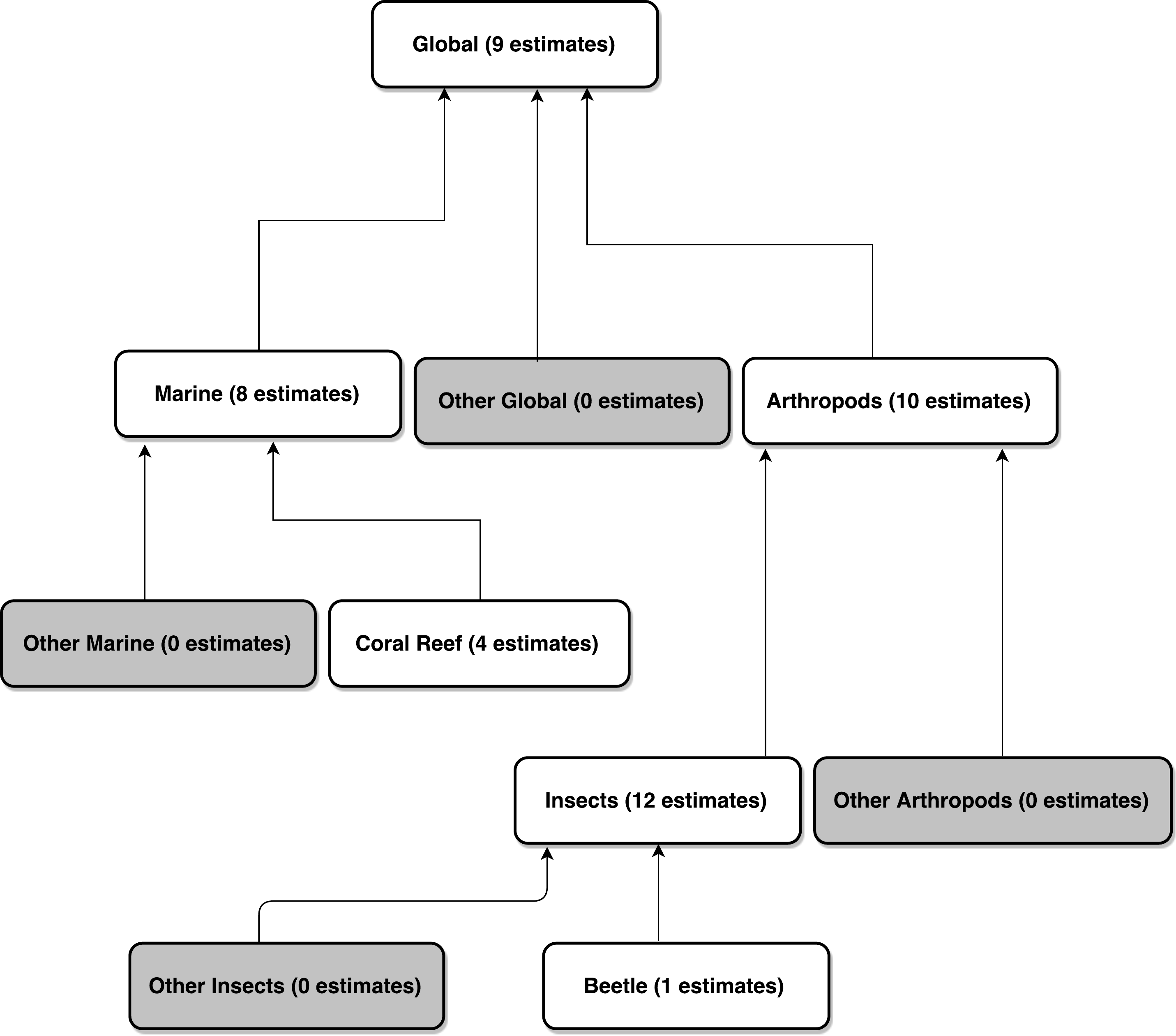}
	\caption{\label{fig:hierarchical}\small Schematic of the hierarchical structure of species categories analysed. White boxes correspond to categories with observed data. Grey boxes correspond to assumed ``other species" categories not observed. 
	}
	\end{figure}
	%
The above model is appropriate for each species category in isolation, however, clearly there is a hierarchical relationship between the different categories. Our observed data consist of global species estimates, coral reefs, marine species, insects and arthropods, and these are naturally related  hierarchically (Figure \ref{fig:hierarchical}). (Note that in Table \ref{Table:dataset} there is a single observation for terrestrial category. However, this datapoint is largely inconsistent with other observations in the insects and arthropods categories, and so it was removed as an observation, and as a species category, from the analysis.) This model indicates that, for example, the number of arthropod species comprises the number of insect species plus the number of other-arthropod species.\\
For some species categories we have observed data and for others none (Figure \ref{fig:hierarchical}, white and grey boxes respectively). In the absence of observed data, the model parameters for these categories can be inferred from the hierarchical structure. For example, the number of arthropod species does not solely consist of the number of insect species. It must also contain a number of other arthropods that have not been estimated individually, or even discovered. Each ``other'' species category is structurally modelled in the same way as the other categories via equation (\ref{eqn:model}), except that there is no observed data. The unknown parameters are therefore determined by the mismatch in estimated parameters from its neighbouring categories. For example, the parameters of the other-arthropods category may be conceptually inferred from the difference between the number of insect species and the number of arthropod species.\\
The model (\ref{eqn:model}) assumes that each data point $(m,\log r)^\top$ is an unbiased estimate of the true species count midpoint ($\mu_{mj}$) and log range ($\mu_{rj}$), and that the interval estimates are exchangeable within each species category (which seems to be supported by \citealt{caley2014global}). If this assumption holds then  the hierarchical structure will allow us to obtain unbiased parameter estimates for the ``other'' species categories.\\
The hierarchical relationship between the species categories in Figure \ref{fig:hierarchical} means that e.g. the sum of the number of beetle and other-insect species should equal the number of insect species, while the sum of the number of insect and other-arthropod species should equal the number of arthropod species, etc. To incorporate this hierarchical structure, we assign the following constraints on the mid-point mean parameter $\mu_m$:
\begin{align*}
&\mu_{m_{insects}}=\mu_{m_{beetles}}+\mu_{m_{other-insects}}\\
&\mu_{m_{arthropods}}=\mu_{m_{insects}}+\mu_{m_{other-arthropods}}\\
&\mu_{m_{marine}}=\mu_{m_{coral reefs}}+\mu_{m_{other-marine}}\\
&\mu_{m_{global}}=\mu_{m_{marine}}+\mu_{m_{arthropods}}	+\mu_{m_{other-global}}.
\end{align*}
Further, it is reasonable to suppose that a similar hierarchical structure also holds for the interval log ranges. By noting that if $X\sim(a_X,b_X)$ and $Y\sim (a_Y,b_Y)$ then $X+Y\sim (a_X+a_Y,b_X+b_Y)$ (where $Z\sim (a_Z,b_Z)$ denotes that the random variable $Z$ is distributed between $a_Z$ and $b_Z$), then, for example, given  that the number of insects must equal the number of beetles plus the number of other insects, then in terms of intervals $(a,b)$ we must have
\begin{eqnarray*}
	a_{insects} & = & a_{beetles} + a_{other-insects}\\
	b_{insects} & = & b_{beetles} + b_{other-insects}.
\end{eqnarray*}
This then implies the constraint 
\begin{align*}
&	\mu_{r_{insects}}  =  \log [2(\mu_{m_{insects}}-\mu_{m_{beetle}}-\mu_{m_{other-insects}})+\exp(\mu_{r_{beetle}})+
\exp(\mu_{r_{other-insects}})]
\end{align*}
on the mean log range parameter for insects, $\mu_{r_{insects}}$. Equivalent constraints on $\mu_r$ also hold for the arthropods and marine categories. The global category, which is the sum of marine, arthropod and other global categories, is similarly constrained
\begin{eqnarray*}
	\mu_{r_{global}} &=&\log[2(\mu_{m_{global}}-\mu_{m_{arthropods}}-\mu_{m_{marine}}-\mu_{m_{other-global}}) \\ &&\qquad\qquad+\exp(\mu_{r_{arthropods}})+ \exp(\mu_{r_{marine}})+\exp(\mu_{r_{other-global}})].
\end{eqnarray*}
In combination, the above constraints mean that the parameters $\mu_j$ and $\Sigma_j$ for the species categories with ``children" categories in Figure \ref{fig:hierarchical} are fully determined by the parameters of their children categories. That is, the only categories with free parameters are beetles, coral reefs and the four ``other" species categories, and once these parameters are known, the parameters for the rest of the hierarchical model are fixed. However, estimation of these parameters accordingly means choosing these parameters so that the observed data could credibly have been observed over the {\em entire} hierarchy, so that e.g. observed data in the global category will directly influence the parameter estimates of all other species categories. In this way our model allows the sharing of information in estimating the number of species in one category given the data in all other categories.\\
Our model is analysed under the Bayesian framework. 
We adopted the following prior specification for the parameters of the non-fixed ``children'' categories:
$\mu_m\sim N(0, 10000)I(\mu_m>0)$ represents a diffuse prior constrained to the positive real line, so that the interval midpoint can be located effectively anywhere.
$\mu_r\sim N(-1,1.5)I(\mu_r<\log(2\mu_m))$ puts most prior weight on smaller ranges, where the constraint ensures that the lower bound of the resulting interval (which depends on both midpoint and range) is always greater than zero.
The standard deviation parameters are specified as $\sigma_m,\sigma_r\sim\mbox{Half-Cauchy}(0,2.5)$ (that is, a Cauchy(0,2.5) distribution constrained to the positive real line), which is a reasonable default choice for a scale parameter in the absence of specific information (\citealt{gelman2006prior}).
Writing $P$ as the correlation matrix associated with each $\Sigma_j$, we specify $P\sim LKJ(1)$ so that the prior for $P$ is uniform over all correlation matrices.
Markov chain Monte Carlo simulation from the resulting posterior distribution was implemented in the Stan software package (\citealt{stan}).


\section*{Results}
\label{results}
\subsection*{Overall Species Estimates}
\label{overall species estimates}
\begin{figure}
	\centering
	\includegraphics[width=13cm]{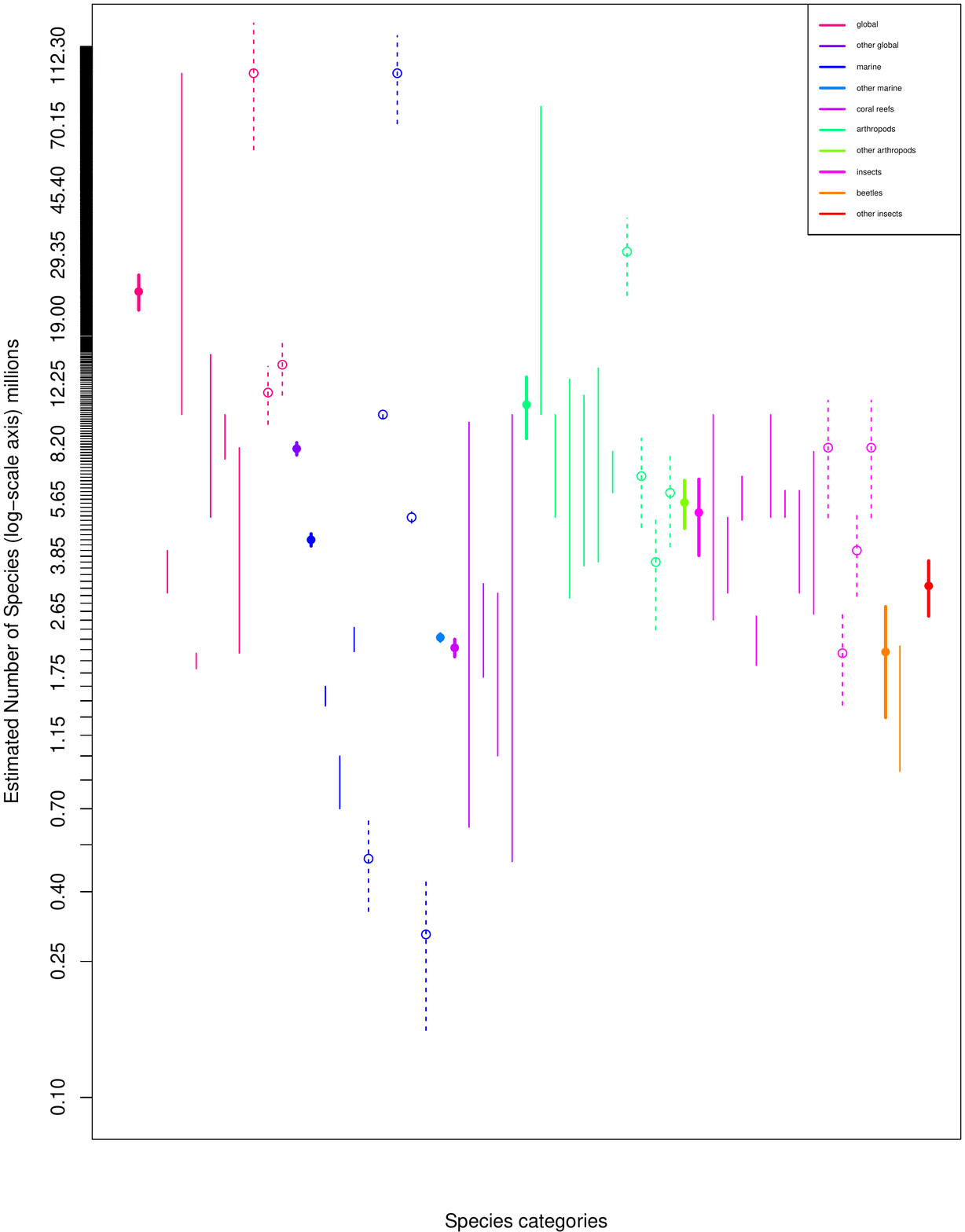}
	\caption{\small Observed data and posterior interval estimates for each species category (as indicated by colour). Open circles and thin lines illustrate observed point ($x$) and interval $(a,b)$ data. Thick lines indicate posterior means of interval for each category, obtained  by inverting the mapping $(m,\log r)^\top\rightarrow(a,b)^\top$ back to the $(a,b)$ parameterisation for the parameters $(\mu_{mj},\mu_{rj})^\top$ of each category. (I.e. we transform the posterior for $(\mu_{mj},\mu_{jr})^\top$ to the posterior for $(\mu_{aj},\mu_{bj})^\top$ where $\mu_{aj}=\mu_{mj}-\exp(\mu_{rj})/2$ and $\mu_{bj}=\mu_{mj}+\exp(\mu_{rj})/2$). The illustrated interval is that obtained from the posterior mean of the lower ($\mu_{aj}$) and upper ($\mu_{bj}$) endpoints of this interval. The filled circle indicates the posterior mean of the interval midpoint $(\mu_{mj})$. Dashed lines indicate posterior predicted posterior mean of interval where only point data $x$ is observed.
	}
	\label{fig:image3}
\end{figure}

Figure \ref{fig:image3} illustrates the observed data and resulting posterior interval summaries from the fitted hierarchical model, plotted on the log scale, for each species category, with different species categories shown by different colours. 
Table \ref{table:post-summary-stats} enumerates some these posterior quantities.

\begin{figure}
	\centering
	\includegraphics[width=10cm]{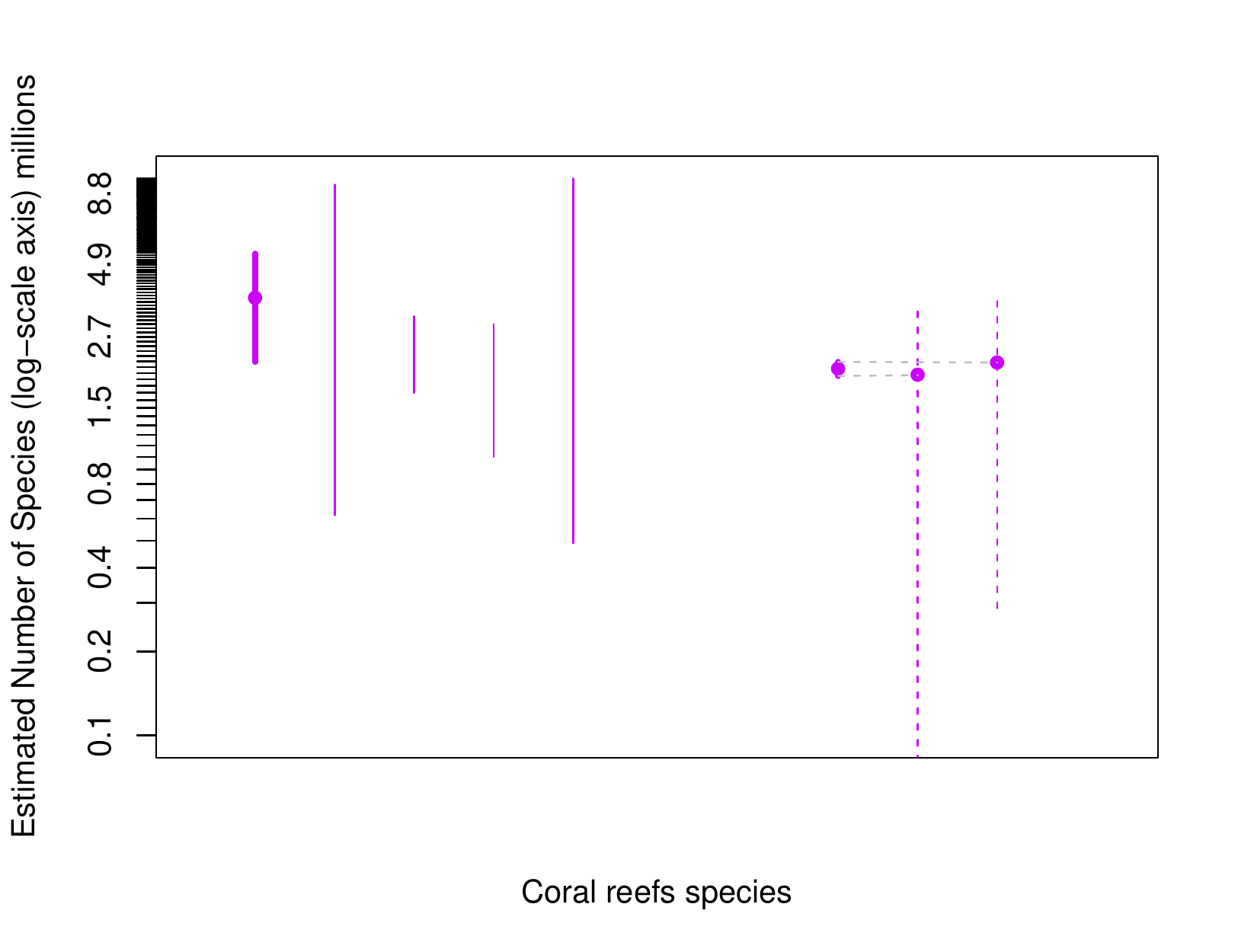}
	\caption{\small As for Figure \ref{fig:image3}, except for a separate analysis of coral reefs with no hierarchical structure (solid lines). The leftmost thick line illustrates the resulting (non-hierarchical) posterior mean interval, whereas the rightmost thick line illustrates the same interval using the full hierarchical model. The two dotted intervals represent 95\% HPD intervals of the lower and upper interval endpoints, based on the full hierarchical model, illustrating considerable uncertainty. (Filled circles represent posterior means.)	}
	\label{fig:check}
\end{figure}

\begin {table}[tb]
\begin{center}
	\begin{tabular}{|l|c|c|c|c|}
		\hline
		& \multicolumn{2}{c|}{Intervals} & \multicolumn{2}{c|}{Midpoints}\\
		Species Category  & Mean lower bound & Mean upper bound &   Mean & 95\% HPD \\ \hline\hline
		Arthropods & 	8.51	 & 	12.87 & 10.69 & (5.09,  17.30) 	 \\ 
		Beetle & 	1.30	 & 	2.74 & 2.02 & (0.40,  4.20) 	 \\ 
		Coral Reefs & 	1.95	 & 	2.19 & 2.07 & (0.16,  3.68) 	 \\ 
		Insects & 	3.87	 & 	6.46 & 5.16 & (3.86,  6.46) 	 \\ 
		Marine & 	4.12	 & 	4.47 & 4.30 & (0.69,  9.04) 	 \\ \hline
		Other Arthropods & 	4.64	 & 	6.41 & 5.53 & (0.08,  11.90) 	 \\ 
		Other Insects  & 	2.57	& 	3.72  & 3.15 & (0.71,  5.07) 	\\ 
		Other Global& 	7.62	 & 	8.27 & 7.94 & (0.03,  19.50) 	 \\ 
		Other Marine& 	2.17	 & 	2.28	& 2.22 & (0.01,  6.71)  \\ \hline
		Global & 	20.25	 & 	25.61 & 22.93 & (11.24, 36.26) 	 \\ \hline
	\end{tabular} 
	\caption{\small Posterior point estimate summaries of species numbers in each category for both intervals and midpoints. Interval estimates are the posterior mean lower and upper interval bound. Midpoint estimates are the posterior mean and the 95\% highest posterior density (HPD) interval.
		Point estimates are measured in millions. 
	}

	\label{table:post-summary-stats}
\end{center}
\end{table}
%
%
The posterior distributions of model parameters incorporate the hierarchical constraints as discussed above (Figure \ref{fig:image3}). This means that, for example, 
the number of beetles (orange lines) plus the number of other insects (red) can be roughly seen to sum to the number of insects (magenta) both as a midpoint (filled circles) and as an interval (note the log scale). Secondly, within each species category the posterior mean intervals (thick lines) and midpoints (filled circles) are mostly consistent with the observed data in each category. This is most clearly seen for insects (magenta), where the length of the posterior mean interval is roughly the average of the observed insect interval lengths, and the posterior  midpoint mean is located roughly at the centre of all the observed point estimates (open circles) and observed interval midpoints. However, this is less apparent for other categories. For example, while the posterior mean midpoint for coral reefs (purple) appears to be well located at the centre of the 4 observed intervals, the posterior mean interval range is considerably shorter than the ranges of the observed data. This is not an error, but is actually a direct and beneficial outcome of the hierarchical model. If  the 4 coral reef interval observations are analysed in isolation using the model (\ref{eqn:model}) and with no hierarchical structure, the resulting posterior mean interval range is more consistent with the ranges of the observed intervals (Figure \ref{fig:check}, as the leftmost thick line).
 
As this simplified model analysis is performing correctly, the difference in outcomes with the full model (rightmost thick line, Figure \ref{fig:check}) must then be due to the hierarchical structure.  This structure states that the interval and midpoint estimates in any category are not just informed by the observed data in that category, but also by the observed data in other categories given their direct and known relationships. In the case of coral reefs, a wide posterior mean interval is just not consistent with the richness estimates that were simultaneously  estimated from the data in other categories (most immediately in marine and other-marine). As a result, given the extra knowledge from the other categories, the model is able to drastically revise its certainty as to credible interval ranges for coral reefs based on the data in related species categories. This would not have been possible without the hierarchical model.\\
In general, the hierarchical analysis is sensible, and preferable to non-hierarchical analyses, both because it allows for a pooling of information over categories which can lead to more precise within-category estimates, and because it can enforce parameter estimation that satisfies known model constraints, thereby automatically generating consistent estimates. An additional benefit of building a symbolic hierarchical model (that is, in combining interval and point estimate data) is that the known positive relationship between interval midpoint and range (Figure \ref{fig:m-versus-logr}) also permits a sharing of information between these two quantities. This means more informed estimates of (say) midpoints are obtained than if a hierarchical model was to be constructed on midpoints alone, which is the standard hierarchical model format. This benefit is in addition to the precision gained by incorporating both interval and point estimate data into the analysis in the first place.\\
One caveat to these benefits is that we are assuming that every observed interval $(a,b)$ or point estimate $x$ is independent and an unbiased estimate of the true quantity for the given category. If this is not the case, then errant observations will not only affect the parameter estimates for their own species category, but they will also influence those in other categories. So there should be a strong emphasis on ensuring data quality and consistency in practice. Note that we excluded the one ``terrestrial" observed point estimate (Table \ref{Table:dataset}; \cite{may1992many}) as it was inconsistent with the hierarchical structure, and to include it would likely negatively affect the remaining category parameter estimates.\\
Finally, in these analyses note that we have only presented the posterior mean interval or posterior mean midpoint as point estimates of these quantities. However in fact there is a full joint posterior distribution associated with them. For example, Table \ref{table:post-summary-stats} presents both the posterior mean and 95\% highest posterior density (HPD) intervals  for the interval midpoints. Note that in many cases these HPD intervals also encompass the mean lower and upper bounds of the associated interval estimates. This is not inconsistent -- the joint posterior distribution for these quantities enforces the constraint $\mu_{aj}\leq\mu_{mj}\leq\mu_{bj}$ absolutely. However, just presenting the posterior marginal mean of each of these parameters hides the fact that there is some uncertainty with each of these parameters, beyond the mean values presented here. For example, the two dashed intervals in Figure \ref{fig:check} illustrate 95\% HPD intervals for the upper and lower interval estimates for coral reefs. The posterior means of these intervals, are then used to construct the posterior mean interval estimates, as indicated by the horizontal dashed lines.


\subsection*{Estimates over time}
\label{time series analysis}

\begin{figure}
	\centering
	\includegraphics[width=13cm]{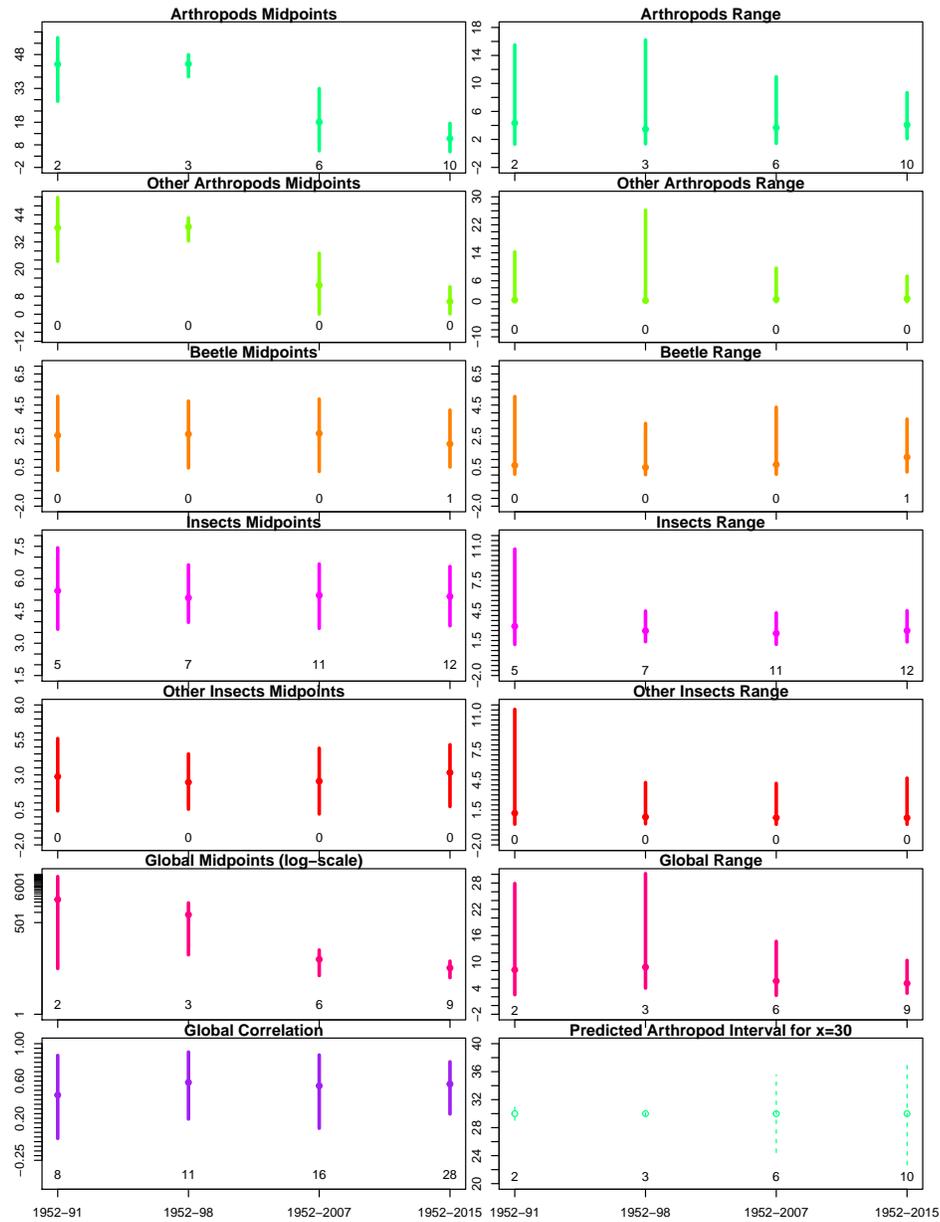}
	\caption{\small Posterior means (filled circles) and 95\% high density credible intervals for interval midpoints ($\mu_{mj}$; left panels) and ranges ($\mu_{rj}$; right panels), estimated from data from four different time periods 1952--1991, 1952--1998, 1952--2007 and 1952--2015. Panels show results for [top to bottom] arthropods, other-arthropods, beetles, insects, other-insects and global, with the number under each graphic indicating the number of directly observed estimates in each category for each time point. 
	The bottom left panel shows  the corresponding correlation between all midpoints and ranges ($\rho$) for each dataset.
	The bottom right panel illustrates the predictive mean interval for the associated observed arthropod point estimate of $x=30$ taken from \cite{erwin1982tropical}.}
	\label{fig:image5_7}
\end{figure}

The final aspect we examine is how the parameters of the hierarchical model evolve as more data are included in the analysis over time. We study this to evaluate how the nature of individual diversity estimates have changed over time, and also to show how the addition of data for one species category under the model affects the species richness estimates for the same or other categories. That there should be some local or global effect is clear due to the nature of the hierarchical model. For example, we might suspect that observing data within one category will have the largest impact on parameter estimates in that category, but there may also be a smaller effect on parameter estimates in other categories. In principle, this should be informative in deciding where best to focus efforts in obtaining future data, to progress towards agreed and more precise estimates of global species richness within or among any species category.\\
In the following we arrange our observed data according to the year the interval or point estimate was published, and construct four (nested) datasets consisting of the data published in the years 1952--1991 (9 observation), 1952--98 (19 observations), 1952--2007 (31 observations) and 1952--2015 (all 44 observations). These year ranges were chosen to include a roughly equal number of new diversity estimates in each successive time period.
We fit our hierarchical symbolic model to the data in each  dataset, and observe how the parameter estimates evolve over time as more data is included in the analysis.
The results are summarised in Figure \ref{fig:image5_7}.

For arthropods, as more data is observed over time, 
the location
and variability of the interval midpoint are reduced substantially. The reduction in variability is expected in the presence of a (relatively) large number of observed datapoints for arthropods (10), but also occurs due to the large amount of information in the neighbouring parent global category (9) and the child category insects (12) (see Figure \ref{fig:hierarchical}), making arthropods one of the most well informed species categories in the hierarchy. 
The reduction in midpoint over time can be primarily attributed to changes from very large early published estimates of arthropod diversity (specifically, $x=30$ from \cite{erwin1982tropical}, and $(a,b)=(10,80)$ from \cite{stork1988insect}), to a consistently lower sequence of eight later published estimates with a mean observed midpoint of $\approx 6.82$ (Table \ref{Table:dataset}). Secondary influences on the midpoint estimate are due to the need for consistency with the rest of the hierarchical model.
%
%
Arthropod interval range estimates also decrease over time, but less dramatically than for the midpoint. In part, this is a result of there being less observed data for ranges than for midpoints for the relevant categories (Figure \ref{fig:image3}: 4/10 arthropod, 3/9 global and 4/12 insect diversity estimates have unobserved ranges). The other-arthropods category is wholly determined by the insects and arthropods categories. As these are both well estimated in the presence of large numbers of observed data, both midpoint and range of the other-arthropods category are particularly well informed, and naturally follow the information within arthropods, despite there being no direct observations in this category. This naturally provides a precise and hierarchically consistent estimate of the likely interval for the diversity of all unobserved arthropods. 

In contrast, a less data-rich section of the hierarchy involves the beetles category (1 observation) along with the parent insects (12 observations) and the unobserved other-insects category (see Figure \ref{fig:hierarchical}). Here, although the estimates for insects become more precise over time, because there is only one observation for beetles in the last time point, there is nothing to distinguish between the beetles and other-insects categories before this datapoint is observed. Hence, for the first three time periods the midpoint and range estimates for beetles and other-insects are highly similar, with their precise values each determined as ``half'' the estimates for insects. Only when the single beetle estimate is included in the final timepoint can some difference be discerned in the beetle midpoints, although with such a small amount of data this still makes differentiating between beetles and other-insects difficult. More direct observations in the beetles category would help resolve this lack of distinction between beetles and other-insects. 
%


As with other species categories, the global diversity estimates clearly get more precise over time as the number of direct estimates in the global category increases, and also as the number of observations at other categories (which determine the global category) also increase. 
%
As with arthropods, the drastic reduction in the location of the global species midpoint is primarily driven by two early and very large global diversity estimates (of $(a,b)=(10,100)$  from \cite{ehrlich1991biodiversity} and $x=100$ from \cite{may1992many}), whereas the subsequent six estimates are more consistent and on a smaller scale. Part of the explanation for the changes in the nature of these estimates (and those in other categories) could arise from an increase in the sophistication of the applied methods. Similarly, in time further species estimate fluctuations could arise in the future with the potential to split species and synonymise currently named species as more genetic data becomes available.

The bottom left  panel illustrates the changes in the global correlation ($\rho$) between interval midpoint ($\mu_{mj}$) and range ($\mu_{rj}$) across all species categories. Clearly as the number of full interval observations $(a,b)$ increases, the correlation between midpoint and range is better estimated. The full dataset analysis correlation is estimated to be moderately strong with a posterior mean of 0.57 and a 95\% HPD interval of (0.25,0.81). 
Finally, the bottom right panel shows the predicted missing ranges of the observed point estimate $x=30$  (million) taken from \cite{erwin1982tropical}. When there is little data, despite there being a positive correlation with the observed interval midpoints, the predicted range 
is strongly influenced by the prior distribution, which places most density on smaller interval ranges.
%
When the arthropod midpoint and range parameters become better estimated, along with their correlation ($\rho$), the predicted interval associated with the $x=30$ point estimate becomes more realistic, and more accurately incorporates information from around the hierarchical model. The final mean predicted interval for this datapoint is (22.7,37.3). 

\section*{Discussion}

Estimates of species richness are typically, although not exclusively, constructed independently within individual realms or taxa. This means that not only are these estimates potentially inconsistent with estimates for other species categories, but there is also an estimation inefficiency in that available information on related taxa is not being accounted for. Here, we have implemented a meta-analysis that addresses both of these issues, while also presenting a technique for bringing both point estimate and interval estimate data forms into the same analysis. In addition, by adopting a hierarchical model, we have also been able to estimate the number of unobserved species in the `other' categories, simply as a result of requiring consistency within the hierarchical model. \\
The outputs of this analysis can be evaluated either in terms of predicted intervals  $(a,b)$ or interval midpoints $(a+b)/2$, if one is prepared to interpret the interval midpoint as a proxy for a point estimate. Under this interpretation, it is 
not
assumed that the distribution within an interval is symmetric, but merely that the midpoint is a convenient parameterisation of the location of the interval, and it must accordingly be interpreted as such. In terms of providing a single point estimate of species diversity, this interpretation must hold at least until more data on asymmetrical bounds become available to permit asymmetric modelling on $[a,x,b]$, the complete interval plus point estimate. If the asymmetrical bounds reported to date in the literature (Table \ref{Table:dataset}) are representative of species categories, the midpoint estimates we provide here are more likely to be over- rather than under-estimates.  This would appear to be less of a problem than current estimates that violate the necessity of species richness categories being a finite partition of global species richness. 

Each of our estimates (Table \ref{table:post-summary-stats}) are broadly consistent with previous estimates in the literature, but with some differences in that they also respect the hierarchical structure of the model and thereby provide a set of estimates that are logically consistent. 
For example, in terrestrial arthropods, for which species richness estimates have begun to narrow (\citealt{stork2015new}), our posterior mean interval of $(8.51, 12.87)$ is wholly above the most recent interval estimate of $(5.9,7.8)$ by \cite{Stork2015}, although our estimate is not inconsistent with the other observed estimates in Table \ref{Table:dataset}. However it is consistent with being the sum of the insects interval estimate of $(3.87,6.46)$ and the other-arthropod estimate of $(4.64,6.41)$. This is not the case for e.g. \cite{Stork2015} who estimate the number of insect species at $(2.6,7.8)$ which has the same upper endpoint as their estimate for arthropods (7.8), implying that there is a positive probability that there are no other species in arthropods apart from insects, or that as a proportion the number of these species is too small to be noticed given current levels of uncertainty.  
Similarly, the global species richness interval estimate of $(7.4, 10)$ by  \cite{mora2011many} (and indeed, most others in Table \ref{Table:dataset}) does not overlap our global posterior mean interval of $(20.25, 25.61)$, determined using currently available species diversity estimates. Again, most reported estimates of global species richness are just not consistent with the estimates in other species categories -- explicitly, they are too low when taking into consideration our hierarchical model.

However, our approach does have some caveats. Primarily, we have assumed that our observed data are independent of each other, and that within any species category the data represent unbiased estimates of the same quantity over time.
The first of these assumptions is unlikely to be true as several of the estimates in Table \ref{Table:dataset} come from the same published study (e.g. \citealt{raven2000nature,novotny2002low,hamilton2010quantifying,costello2011predicting,Stork2015} and each provide multiple estimates).
The second assumption is also unlikely to be true as knowledge has increased since the global species richness estimate of $(3,4)$ by \cite{raven1983challenge}, definitions of which species are in which category have changed, and the number of actual species has itself changed through the wide spread application of molecular genetic analyses with the power to split cryptic species and synonymize morphospecies. Together this implies there are potential quality issues with these data and the methods used to analyse them that need to be acknowledged. In our opinion, it should no longer be acceptable to simply claim primacy of some method of estimation of global species richness over all others without first ensuring that the method does not violate the finite partition of global species richness within and beyond the taxa of interest, nor without presenting some form of validation of the degree to which the method provides more accurate and precise estimates. Opportunities, however, for such validation will be limited but may in some circumstances be supported by the application of expert knowledge (\cite{fisher2015species}). Superior logic could be argued, for example, if an estimation method incorporates more realistic assumptions and knowledge about how species are distributed in space that were absent from previous methods. Such knowledge of these spatial distributions could also make available information from numerous estimates of species richness at sub-global scales, but doing so will require additional model complexity to accommodate estimates of beta-diversity. Whatever the case, the implications of these choices on the results obtained should be clearly presented. Where these things can be accomplished, it may be justified to weight the contributions to the likelihood of each observed data point or interval, according to the explicit justification of its reliability. Fully reliable observations would be weighted 1, completely unreliable observations 0, and so would be  effectively removed from the analysis, as we implemented here with May’s 1992 estimate of the number of terrestrial species. Multiple dependent observations from the same study would receive a weight between these extremes. Ultimately, however, true validation and confidence in these estimates will only be available over the longer-term as the discovery of new species and their taxonomy and systematics proceeds. This progress will provide the opportunity to adaptively learn from testing new estimates against old to assess progress toward convergence as these new estimates propagate up through the hierarchy, and thereby, facilitating the exploration of the consequences of these new estimates on estimates of global species richness.

\newpage
\bibliography{biodiversity}

\begin{thebibliography}{}

\bibitem[\protect\citeauthoryear{Appeltans, Ahyong, Anderson, Angel, Artois,
  Bailly, Bamber, Barber, Bartsch, Berta, et~al.}{Appeltans
  et~al.}{2012}]{appeltans2012magnitude}
Appeltans, W., S.~T. Ahyong, G.~Anderson, M.~V. Angel, T.~Artois, N.~Bailly,
  R.~Bamber, A.~Barber, I.~Bartsch, A.~Berta, et~al. (2012).
\newblock The magnitude of global marine species diversity.
\newblock {\em Current Biology\/}~{\em 22\/}(23), 2189--2202.

\bibitem[\protect\citeauthoryear{Basset, Samuelson, Allison, and Miller}{Basset
  et~al.}{1996}]{basset1996many}
Basset, Y., G.~Samuelson, A.~Allison, and S.~Miller (1996).
\newblock How many species of host-specific insects feed on a species of
  tropical tree?
\newblock {\em Biological Journal of the Linnean Society\/}~{\em 59\/}(2),
  201--216.

\bibitem[\protect\citeauthoryear{Benayas, Newton, Diaz, and Bullock}{Benayas
  et~al.}{2009}]{benayas2009enhancement}
Benayas, J. M.~R., A.~C. Newton, A.~Diaz, and J.~M. Bullock (2009).
\newblock Enhancement of biodiversity and ecosystem services by ecological
  restoration: a meta-analysis.
\newblock {\em Science\/}~{\em 325\/}(5944), 1121--1124.

\bibitem[\protect\citeauthoryear{Billard and Diday}{Billard and
  Diday}{2007}]{billarddiday2007}
Billard, L. and E.~Diday (2007).
\newblock {\em Symbolic Data Analysis : Conceptual Statistics and Data Mining}.
\newblock John Wiley and Sons Ltd.

\bibitem[\protect\citeauthoryear{Bouchet and Duarte}{Bouchet and
  Duarte}{2006}]{bouchet2006exploration}
Bouchet, P. and C.~M. Duarte (2006).
\newblock The exploration of marine biodiversity: scientific and technological
  challenges.
\newblock {\em Fundaci{\'o}n BBVA\/}~{\em 33}.

\bibitem[\protect\citeauthoryear{Brito and Duarte~Silva}{Brito and
  Duarte~Silva}{2012}]{brito2012modelling}
Brito, P. and A.~P. Duarte~Silva (2012).
\newblock Modelling interval data with normal and skew-normal distributions.
\newblock {\em Journal of Applied Statistics\/}~{\em 39\/}(1), 3--20.

\bibitem[\protect\citeauthoryear{Caley, Fisher, and Mengersen}{Caley
  et~al.}{2014}]{caley2014global}
Caley, M.~J., R.~Fisher, and K.~Mengersen (2014).
\newblock Global species richness estimates have not converged.
\newblock {\em Trends in Ecology \& Evolution\/}~{\em 29\/}(4), 187--188.

\bibitem[\protect\citeauthoryear{Carpenter, Gelman, Hoffman, Lee, Goodrich,
  Betancourt, Brubaker, Guo, Li, and Riddell}{Carpenter et~al.}{2015}]{stan}
Carpenter, B., A.~Gelman, M.~Hoffman, D.~Lee, B.~Goodrich, M.~Betancourt,
  M.~Brubaker, J.~Guo, P.~Li, and A.~Riddell (2015).
\newblock Stan: {A} probabilistic programming language.
\newblock {\em Journal of Statistical Software\/}, in press.

\bibitem[\protect\citeauthoryear{Ceballos, Ehrlich, Barnosky, Garc{\'\i}a,
  Pringle, and Palmer}{Ceballos et~al.}{2015}]{ceballos2015accelerated}
Ceballos, G., P.~R. Ehrlich, A.~D. Barnosky, A.~Garc{\'\i}a, R.~M. Pringle, and
  T.~M. Palmer (2015).
\newblock Accelerated modern human--induced species losses: Entering the sixth
  mass extinction.
\newblock {\em Science Advances\/}~{\em 1\/}(5), e1400253.

\bibitem[\protect\citeauthoryear{Costello, May, and Stork}{Costello
  et~al.}{2013}]{costello2013can}
Costello, M.~J., R.~M. May, and N.~E. Stork (2013).
\newblock Can we name earth's species before they go extinct?
\newblock {\em Science\/}~{\em 339\/}(6118), 413--416.

\bibitem[\protect\citeauthoryear{Costello, Wilson, and Houlding}{Costello
  et~al.}{2011}]{costello2011predicting}
Costello, M.~J., S.~Wilson, and B.~Houlding (2011).
\newblock Predicting total global species richness using rates of species
  description and estimates of taxonomic effort.
\newblock {\em Systematic Biology\/}, syr080.

\bibitem[\protect\citeauthoryear{Cracraft and Grifo}{Cracraft and
  Grifo}{1999}]{cracraft1999living}
Cracraft, J. and F.~T. Grifo (1999).
\newblock {\em The living planet in crisis: biodiversity science and policy}.
\newblock Columbia University Press.

\bibitem[\protect\citeauthoryear{Ehrlich and Wilson}{Ehrlich and
  Wilson}{1991}]{ehrlich1991biodiversity}
Ehrlich, P.~R. and E.~O. Wilson (1991).
\newblock Biodiversity studies: science and policy.
\newblock {\em Science\/}~{\em 253\/}(5021), 758.

\bibitem[\protect\citeauthoryear{Erwin}{Erwin}{1982}]{erwin1982tropical}
Erwin, T.~L. (1982).
\newblock Tropical forests: their richness in coleoptera and other arthropod
  species.
\newblock {\em Coleopterists Bulletin\/}~{\em 36\/}(1), 74--75.

\bibitem[\protect\citeauthoryear{Fisher, O'Leary, Low-Choy, Mengersen,
  Knowlton, Brainard, and Caley}{Fisher et~al.}{2015}]{fisher2015species}
Fisher, R., R.~A. O'Leary, S.~Low-Choy, K.~Mengersen, N.~Knowlton, R.~E.
  Brainard, and M.~J. Caley (2015).
\newblock Species richness on coral reefs and the pursuit of convergent global
  estimates.
\newblock {\em Current Biology\/}~{\em 25\/}(4), 500--505.

\bibitem[\protect\citeauthoryear{Gaston}{Gaston}{1991}]{gaston1991magnitude}
Gaston, K.~J. (1991).
\newblock The magnitude of global insect species richness.
\newblock {\em Conservation biology\/}~{\em 5\/}(3), 283--296.

\bibitem[\protect\citeauthoryear{Gelman et~al.}{Gelman
  et~al.}{2006}]{gelman2006prior}
Gelman, A. et~al. (2006).
\newblock Prior distributions for variance parameters in hierarchical models
  {(Comment on a paper by Browne and Draper)}.
\newblock {\em Bayesian Analysis\/}~{\em 1\/}(3), 515--534.

\bibitem[\protect\citeauthoryear{Gelman, Carlin, and Stern}{Gelman
  et~al.}{2003}]{gelmanbook}
Gelman, A., J.~B. Carlin, and H.~S. Stern (2003).
\newblock {\em Bayesian Data Analysis}.
\newblock Chapman \& Hall/CRC Press.

\bibitem[\protect\citeauthoryear{Grassle and Maciolek}{Grassle and
  Maciolek}{1992}]{grassle1992deep}
Grassle, J.~F. and N.~J. Maciolek (1992).
\newblock Deep-sea species richness: {Regional} and local diversity estimates
  from quantitative bottom samples.
\newblock {\em American Naturalist\/}, 313--341.

\bibitem[\protect\citeauthoryear{Groombridge and Jenkins}{Groombridge and
  Jenkins}{2002}]{groombridge2002world}
Groombridge, B. and M.~D. Jenkins (2002).
\newblock World atlas of biodiversity.
\newblock {\em Prepared by the UNEP World Conservation Monitoring Centre.
  Unversity of California Press, Berkeley\/}.

\bibitem[\protect\citeauthoryear{Hamilton, Basset, Benke, Grimbacher, Miller,
  Novotn{\`y}, Samuelson, Stork, Weiblen, and Yen}{Hamilton
  et~al.}{2010}]{hamilton2010quantifying}
Hamilton, A.~J., Y.~Basset, K.~K. Benke, P.~S. Grimbacher, S.~E. Miller,
  V.~Novotn{\`y}, G.~A. Samuelson, N.~E. Stork, G.~D. Weiblen, and J.~D. Yen
  (2010).
\newblock Quantifying uncertainty in estimation of tropical arthropod species
  richness.
\newblock {\em The American Naturalist\/}~{\em 176\/}(1), 90--95.

\bibitem[\protect\citeauthoryear{Hamilton, Novotn{\`y}, Waters, Basset, Benke,
  Grimbacher, Miller, Samuelson, Weiblen, Yen, et~al.}{Hamilton
  et~al.}{2013}]{hamilton2013estimating}
Hamilton, A.~J., V.~Novotn{\`y}, E.~K. Waters, Y.~Basset, K.~K. Benke, P.~S.
  Grimbacher, S.~E. Miller, G.~A. Samuelson, G.~D. Weiblen, J.~D. Yen, et~al.
  (2013).
\newblock Estimating global arthropod species richness: refining probabilistic
  models using probability bounds analysis.
\newblock {\em Oecologia\/}~{\em 171\/}(2), 357--365.

\bibitem[\protect\citeauthoryear{Hammond}{Hammond}{1995}]{hammond1995described}
Hammond, P. (1995).
\newblock Described and estimated species numbers: an objective assessment of
  current knowledge.
\newblock {\em Microbial diversity and ecosystem function\/}, 29--71.

\bibitem[\protect\citeauthoryear{Hodkinson and Casson}{Hodkinson and
  Casson}{1991}]{hodkinson1991lesser}
Hodkinson, I. and D.~Casson (1991).
\newblock A lesser predilection for bugs: Hemiptera (insecta) diversity in
  tropical rain forests.
\newblock {\em Biological Journal of the Linnean Society\/}~{\em 43\/}(2),
  101--109.

\bibitem[\protect\citeauthoryear{Hunter}{Hunter}{2007}]{hunter2007human}
Hunter, P. (2007).
\newblock The human impact on biological diversity.
\newblock {\em EMBO reports\/}~{\em 8\/}(4), 316--318.

\bibitem[\protect\citeauthoryear{Knowlton, Brainard, Fisher, Moews, Plaisance,
  and Caley}{Knowlton et~al.}{2010}]{knowlton2010coral}
Knowlton, N., R.~E. Brainard, R.~Fisher, M.~Moews, L.~Plaisance, and M.~J.
  Caley (2010).
\newblock Coral reef biodiversity.
\newblock {\em Life in the Worlds Oceans: Diversity Distribution and
  Abundance\/}, 65--74.

\bibitem[\protect\citeauthoryear{Lambshead}{Lambshead}{1993}]{lambshead1993recent}
Lambshead, P. (1993).
\newblock Recent developments in marine benthic biodiversity research.
\newblock {\em Oceanis\/}~{\em 19}, 5--5.

\bibitem[\protect\citeauthoryear{Le-Rademacher and Billard}{Le-Rademacher and
  Billard}{2011}]{le2011likelihood}
Le-Rademacher, J. and L.~Billard (2011).
\newblock Likelihood functions and some maximum likelihood estimators for
  symbolic data.
\newblock {\em Journal of Statistical Planning and Inference\/}~{\em 141\/}(4),
  1593--1602.

\bibitem[\protect\citeauthoryear{May}{May}{1992a}]{may1992bottoms}
May, R.~M. (1992a).
\newblock Bottoms up for the oceans.
\newblock {\em Nature\/}~{\em 357}, 278--279.

\bibitem[\protect\citeauthoryear{May}{May}{1992b}]{may1992many}
May, R.~M. (1992b).
\newblock How many species inhabit the earth.
\newblock {\em Scientific American\/}~{\em 267\/}(4), 42--48.

\bibitem[\protect\citeauthoryear{May and Beverton}{May and
  Beverton}{1990}]{may1990many}
May, R.~M. and R.~Beverton (1990).
\newblock How many species?[and discussion].
\newblock {\em Philosophical Transactions of the Royal Society of London B:
  Biological Sciences\/}~{\em 330\/}(1257), 293--304.

\bibitem[\protect\citeauthoryear{Mooney, Larigauderie, Cesario, Elmquist,
  Hoegh-Guldberg, Lavorel, Mace, Palmer, Scholes, and Yahara}{Mooney
  et~al.}{2009}]{mooney2009biodiversity}
Mooney, H., A.~Larigauderie, M.~Cesario, T.~Elmquist, O.~Hoegh-Guldberg,
  S.~Lavorel, G.~M. Mace, M.~Palmer, R.~Scholes, and T.~Yahara (2009).
\newblock Biodiversity, climate change, and ecosystem services.
\newblock {\em Current Opinion in Environmental Sustainability\/}~{\em 1\/}(1),
  46--54.

\bibitem[\protect\citeauthoryear{Mora, Tittensor, Adl, Simpson, and Worm}{Mora
  et~al.}{2011}]{mora2011many}
Mora, C., D.~P. Tittensor, S.~Adl, A.~G. Simpson, and B.~Worm (2011).
\newblock How many species are there on earth and in the ocean?
\newblock {\em PLoS Biol\/}~{\em 9\/}(8), e1001127.

\bibitem[\protect\citeauthoryear{Nielsen and Mound}{Nielsen and
  Mound}{2000}]{nielsen2000global}
Nielsen, E.~S. and L.~A. Mound (2000).
\newblock Global diversity of insects: the problems of estimating numbers.
\newblock {\em Nature and human society: The quest for a sustainable world\/},
  213--222.

\bibitem[\protect\citeauthoryear{Novotny, Basset, Miller, Weiblen, Bremer,
  Cizek, and Drozd}{Novotny et~al.}{2002}]{novotny2002low}
Novotny, V., Y.~Basset, S.~E. Miller, G.~D. Weiblen, B.~Bremer, L.~Cizek, and
  P.~Drozd (2002).
\newblock Low host specificity of herbivorous insects in a tropical forest.
\newblock {\em Nature\/}~{\em 416\/}(6883), 841--844.

\bibitem[\protect\citeauthoryear{{\O}Degaard}{{\O}Degaard}{2000}]{odegaard2000}
{\O}Degaard, F. (2000).
\newblock How many species of arthropods? {Erwin's} estimate revised.
\newblock {\em Biological Journal of the Linnean Society\/}~{\em 71\/}(4),
  583--597.

\bibitem[\protect\citeauthoryear{Pimm, Jenkins, Abell, Brooks, Gittleman,
  Joppa, Raven, Roberts, and Sexton}{Pimm et~al.}{2014}]{pimm2014biodiversity}
Pimm, S.~L., C.~N. Jenkins, R.~Abell, T.~M. Brooks, J.~L. Gittleman, L.~N.
  Joppa, P.~H. Raven, C.~M. Roberts, and J.~O. Sexton (2014).
\newblock The biodiversity of species and their rates of extinction,
  distribution, and protection.
\newblock {\em Science\/}~{\em 344\/}(6187), 1246752.

\bibitem[\protect\citeauthoryear{Raven}{Raven}{1983}]{raven1983challenge}
Raven, P.~H. (1983).
\newblock The challenge of tropical biology.
\newblock {\em Bulletin of the Entomological Society of America\/}~{\em
  29\/}(1), 4--13.

\bibitem[\protect\citeauthoryear{Raven et~al.}{Raven
  et~al.}{2000}]{raven2000nature}
Raven, P.~H. et~al. (2000).
\newblock {\em Nature and human society: the quest for a sustainable world}.
\newblock National Academies.

\bibitem[\protect\citeauthoryear{Raven and Yeates}{Raven and
  Yeates}{2007}]{raven2007australian}
Raven, P.~H. and D.~K. Yeates (2007).
\newblock Australian biodiversity: threats for the present, opportunities for
  the future.
\newblock {\em Australian Journal of Entomology\/}~{\em 46\/}(3), 177--187.

\bibitem[\protect\citeauthoryear{Reaka-Kudla}{Reaka-Kudla}{2005}]{reaka2005}
Reaka-Kudla (2005).
\newblock {\em Biodiversity of Caribbean coral reefs. In: Caribbean Marine
  Biodiversity: The Known and the Unknown.}
\newblock DEStech Publications.

\bibitem[\protect\citeauthoryear{Reaka-Kudla, Wilson, and Wilson}{Reaka-Kudla
  et~al.}{1996}]{reaka1996biodiversity}
Reaka-Kudla, M.~L., D.~E. Wilson, and E.~O. Wilson (1996).
\newblock {\em Biodiversity II: understanding and protecting our biological
  resources}.
\newblock Joseph Henry Press.

\bibitem[\protect\citeauthoryear{Sabrosky}{Sabrosky}{1953}]{sabrosky1953many}
Sabrosky, C.~W. (1953).
\newblock How many insects are there?
\newblock {\em Systematic Zoology\/}~{\em 2\/}(1), 31--36.

\bibitem[\protect\citeauthoryear{Small, Adey, and Spoon}{Small
  et~al.}{1998}]{small1998current}
Small, A.~M., W.~H. Adey, and D.~Spoon (1998).
\newblock Are current estimates of coral reef biodiversity too low? {The} view
  through the window of a microcosm.
\newblock {\em Atoll Research Bulletin\/}~{\em 450}, 20.

\bibitem[\protect\citeauthoryear{Stork and Gaston}{Stork and
  Gaston}{1990}]{stork1990counting}
Stork, N. and K.~Gaston (1990).
\newblock Counting species one by one.
\newblock {\em New Scientist\/}~{\em 127\/}(1729), 43--47.

\bibitem[\protect\citeauthoryear{Stork}{Stork}{1988}]{stork1988insect}
Stork, N.~E. (1988).
\newblock Insect diversity: facts, fiction and speculation.
\newblock {\em Biological Journal of the Linnean Society\/}~{\em 35\/}(4),
  321--337.

\bibitem[\protect\citeauthoryear{Stork, McBroom, Gely, and Hamilton}{Stork
  et~al.}{2015a}]{stork2015new}
Stork, N.~E., J.~McBroom, C.~Gely, and A.~J. Hamilton (2015a).
\newblock New approaches narrow global species estimates for beetles, insects,
  and terrestrial arthropods.
\newblock {\em Proceedings of the National Academy of Sciences\/}~{\em
  112\/}(24), 7519--7523.

\bibitem[\protect\citeauthoryear{Stork, McBroom, Gely, and Hamilton}{Stork
  et~al.}{2015b}]{Stork2015}
Stork, N.~E., J.~McBroom, C.~Gely, and A.~J. Hamilton (2015b).
\newblock New approaches narrow global species estimates for beetles, insects,
  and terrestrial arthropods.
\newblock {\em Proceedings of the National Academy of Sciences\/}~{\em
  112\/}(24), 7519--7523.

\bibitem[\protect\citeauthoryear{Zhang and Sisson}{Zhang and
  Sisson}{2016}]{zhang+s16}
Zhang, X. and S.~A. Sisson (2016).
\newblock Constructing likelihood functions for interval valued data.
\newblock {\em arXiv:1608.00107\/}.

\end{thebibliography}



\renewcommand{\thetable}{\arabic{table}}
\begin{table}[h]
\singlespacing
	\centering
	\begin{tabular}{|l|l|l|l|l|}
		\hline
		Species Category  & 	  Lower ($a$)	 &   Point estimate ($x$) & 	 Upper  ($b$) & 	Source\\ \hline\hline
		Beetles & 	1.5	& & 	2.1& 	\cite{stork2015new}*\\\hline
		Coral Reefs  & 	0.62	 &  & 	9.5	 & \cite{reaka1996biodiversity}\\ 
		Coral Reefs & 	1.7	 &  & 	3.2	 & \cite{small1998current}\\ 
		Coral Reefs& 	1& 	2	& 3& 	\cite{reaka2005}\\ 
		Coral Reefs& 	0.49& & 		10& 	\cite{knowlton2010coral} \\ \hline
		Global & 	3	& & 	4& 	\cite{raven1983challenge}\\ 
		Global  & 	10	& & 	100	& \cite{ehrlich1991biodiversity}\\ 
		Global & 	& 	100	& &  	\cite{may1992many}\\
		Global & & 	  	11.6& & 	 	\cite{cracraft1999living}\\
		Global & 	5& 	7$^a$& 	15& 	\cite{raven2000nature}\\
		Global & & 		14	& &  	\cite{groombridge2002world}\\
		Global & 	7.4& 	8.7	& 10& 	\cite{mora2011many}\\
		Global & 	1.8	& & 	2& 	\cite{costello2011predicting}\\
		Global & 	2& 	5	& 8	& \cite{costello2013can}\\\hline
		Marine 	& & 	5	 & & 	\cite{may1990many}\\
		Marine  & & 		10	& &  	\cite{grassle1992deep}\\
		Marine  	& & 	100	& &  	\cite{lambshead1993recent}\\
		Marine   	& & 	0.5& & 	 	\cite{raven2000nature}\\
		Marine & 	1.4	& & 	1.6& 	\cite{bouchet2006exploration}\\
		Marine & 	2.02& 	2.2& 	2.38& 	\cite{mora2011many}\\
		Marine   & 	& 	0.3	& &  	\cite{costello2011predicting}\\
		Marine& 	0.7	& & 	1& 	\cite{appeltans2012magnitude}\\\hline
		Terrestrial 	  & &  	10	 & & 	\cite{may1992many}$\dagger$\\\hline
		Terrestrial (Arthropods)& & 	  	30	& &  	\cite{erwin1982tropical}\\
		Terrestrial (Arthropods)& 	10	& & 	80& 	\cite{stork1988insect}\\
		Terrestrial (Arthropods)	 & &   	6.6	 & & 	\cite{basset1996many}\\
		Terrestrial (Arthropods)& 	5	& & 	10	& \cite{odegaard2000}\\
		Terrestrial (Arthropods)	& & 	3.7	 & & 	\cite{novotny2002low}\\
		Terrestrial (Arthropods) 	& & 	5.9	 & & 	\cite{novotny2002low}\\
		Terrestrial (Arthropods)& 	3.6& 	6.1$^a$	& 11.4& 	\cite{hamilton2010quantifying}\\
		Terrestrial (Arthropods)& 	3.7	& 7.8$^a$& 	13.7& 	\cite{hamilton2010quantifying}\\
		Terrestrial (Arthropods)& 	2.9	& & 	12.7& 	\cite{hamilton2013estimating}\\
		Terrestrial (Arthropods)& 	5.9	& 6.8$^a$& 	7.8& 	\cite{stork2015new}*\\\hline
		Terrestrial (Insects) & 	2.5& & 		10	& \cite{sabrosky1953many} \\
		Terrestrial (Insects)& 	3& & 		5& 	\cite{may1990many}\\
		Terrestrial (Insects) & 	4.9	& & 	6.6	& \cite{stork1990counting}\\
		Terrestrial (Insects)  & 	1.84	& & 	2.57& 	\cite{hodkinson1991lesser}\\
		Terrestrial (Insects)&  	5	& & 	10& 	\cite{gaston1991magnitude}\\
		Terrestrial (Insects)  & & 		8& & 	 	\cite{hammond1995described}\\
		Terrestrial (Insects) & 	3	& & 	6& 	\cite{raven2000nature}\\
		Terrestrial (Insects)   	& & 	4	& &   	\cite{raven2000nature}\\
		Terrestrial (Insects)   	& & 	2	& &   	\cite{nielsen2000global}\\
		Terrestrial (Insects)   	& & 	8	& &   	\cite{groombridge2002world}\\
		Terrestrial (Insects) & 	5	& & 	6& 	\cite{raven2007australian}\\
		Terrestrial (Insects)& 	2.6	&5.5$^{a}$ & 	7.8& 	\cite{stork2015new}*\\\hline
		\end{tabular}
\caption{\label{Table:dataset}\small Point ($x$) and interval $(a,b)$ estimates of species diversity from 45 previously published studies. 
Diversity estimates are measured in millions. These data were originally collated by \cite{caley2014global} with the exception of those in \cite{stork2015new}, as indicated by asterisks *. $\dagger$ indicates that this datapoint was not used in this analysis as it is strongly inconsistent with all other estimates. $^{a}$ indicates that the point estimate is asymmetric with respect to the interval, so that $x\neq(a+b)/2$.}
\end{table}

\newpage{}

\end{document}